\documentclass[proceedings]{JHEP3}

\PrHEP{PrHEP hep2001}                   
\conference{International Europhysics Conference on HEP}                

\def\refse#1{\mbox{Sect.~\ref{#1}}}
\def\refses#1{\mbox{Sects.~\ref{#1}}}

\def\citere#1{\mbox{Ref.~\cite{#1}}}
\def\citeres#1{\mbox{Refs.~\cite{#1}}}

\def\al{\alpha}
\def\be{\beta}
\def\ga{\gamma}

\def\si{\sigma}

\newcommand{\TeV}{\unskip\,\mathrm{TeV}}

\def\mathswitchr#1{\relax\ifmmode{\mathrm{#1}}\else$\mathrm{#1}$\fi}

\newcommand{\PW}{\mathswitchr W}
\newcommand{\PZ}{\mathswitchr Z}

\newcommand{\Pne}{\mathswitch \nu_{\mathrm{e}}}

\newcommand{\Pf}{\mathswitch f}

\newcommand{\Pq}{\mathswitchr q}
\newcommand{\Pep}{\mathswitchr {e^+}}
\newcommand{\Pem}{\mathswitchr {e^-}}

\def\mathswitch#1{\relax\ifmmode#1\else$#1$\fi}

\newcommand{\MW}{\mathswitch {M_\PW}}

\def\ie{i.e.\ }

\def\etal{{\it et al.}}

\newcommand{\SU}{\mathrm{SU}}

\newcommand{\SUtwo}{\mathrm{SU(2)}}
\newcommand{\Uone}{\mathrm{U}(1)}

\title{Electroweak Radiative Corrections at High Energies}%
\author{
        Ansgar Denner%
        \thanks{This work was supported in part by the Swiss Bundesamt
          f\"ur Bildung und Wissenschaft and by the
          European Union under contract HPRN-CT-2000-00149.}
\\
        Paul Scherrer Institut, 5232 Villigen PSI,
        Switzerland\\                                        
        E-mail: \email{Ansgar.Denner@psi.ch}}                      

      \abstract{For energies far above the electroweak scale,
        large electroweak radiative corrections occur that grow
        logarithmically with energy and can easily reach several tens
        of per cent in the TeV range. Recent work on these corrections
        is reviewed.}

\begin{document}

\section{Introduction}
  
In the energy range above the electroweak scale, $\sqrt{s}\gg \MW$,
large electroweak radiative corrections occur, which are due to
logarithms $\log{(s/\MW^2)}$ involving the ratio of the energy to the
electroweak scale
\cite{Kuroda:1991wn,Beenakker:1993tt,Ciafaloni:1999xg}.  Such
corrections grow with energy, and at $\sqrt{s}=0.5$--$1\TeV$ they are
typically of order $10\%$ of the theoretical prediction.

For electroweak processes that are not mass-suppressed at high
energies, these logarithmic corrections are universal.  On the
one hand, single logarithms originating from short-distance scales
are related to the renormalization of dimensionless parameters, \ie the
running of the gauge, Yukawa, and scalar couplings. These are governed
by the renormalization group. On the other hand, based on our
experience from QCD, universal logarithms originating from the
long-distance scale $\MW\ll\sqrt{s}$ are expected to factorize, \ie
they can be associated with external lines
in Feynman diagrams.  At the one-loop level, they consist of
double-logarithmic and single-logarithmic terms originating
from soft--collinear and collinear (or soft) gauge bosons,
respectively, coupled to external legs. The non-logarithmic terms, on
the other hand, are in general non-universal and have to be evaluated
\looseness -1
for each process separately, if needed.

In the recent literature most interest has been devoted to electroweak
long-distance corrections.  The main difference between QCD and the
Electroweak Standard Model is that the masses of the weak gauge bosons
provide a physical cut-off for real $\PZ$- and $\PW$-boson emission.
Therefore, for a sufficiently good experimental resolution, soft and
collinear weak-boson radiation need not be included in the theoretical
predictions and, except for electromagnetic real corrections, one can
restrict oneself to large logarithms originating from virtual
corrections. This is assumed in
\refses{se:one-loop}--\ref{se:resummation}, if not stated otherwise.
 
\section{Explicit one-loop calculations}
\label{se:one-loop}

One approach to the logarithmic electroweak corrections is via
explicit calculation from Feynman diagrams. This approach has been
mainly applied at the one-loop level and often starts from complete
calculations of the one-loop corrections. As compared to the complete
results, the results in logarithmic approximation are shorter and allow
analytical studies of the structure of the corrections.

The first evaluation of electroweak corrections at high energies was
already performed many years ago by Beenakker \etal\ \cite{Beenakker:1993tt}.
In this work, the electroweak corrections to on-shell $\PW$-pair
production were evaluated in the high-energy limit including besides
the logarithmic contributions also the constant terms. 

More recently, the virtual logarithmic corrections to
$\Pep\Pem\to\Pf\bar{\Pf}$ have been considered in a series of papers
using explicit high-energy expansions for Feynman diagrams. Ciafaloni
and Comelli \cite{Ciafaloni:1999xg} have pointed out the role of the
Sudakov double logarithms and discussed their origin.  Beccaria \etal\ 
have considered the complete logarithmic corrections and studied their
impact on various observables for light fermions
\cite{Beccaria:2000fk}, for bottom quarks \cite{Beccaria:2000xd}, and
for top quarks \cite{Beccaria:2001jz} in the final state. In
\citeres{Beccaria:2001jz,Beccaria:2001vb} the single logarithmic
corrections appearing in the Minimal Supersymmetric Standard Model
(MSSM) were included.  In all these papers the gauge-invariant subset
of electromagnetic corrections has been split off, and only the weak
corrections have been analysed. In \citere{Beccaria:2001yf} the weak
logarithmic corrections have been separated into universal,
angular-dependent, and Yukawa contributions, and the numerical
importance of angular-dependent corrections was emphasized.  While the
total one-loop contribution remains at the level of a few per cent at
$3\TeV$, individual contributions reach 10\%.  In
\citere{Beccaria:2001an}, it was shown that the slopes of a number of
observables at energies around $3\TeV$ depend only on $\tan\be$, thus
potentially allowing to measure this parameter of the MSSM at CLIC for
$\tan\be<2$ and $\tan\be>20$ with acceptable precision.

Layssac \etal\ \cite{Layssac:2001ur} have evaluated the complete
logarithmic corrections for the process $\ga\ga\to\Pf\bar\Pf$
including the case of heavy-quark production in the Standard Model and
the MSSM. Also in this case the gauge-invariant QED corrections have
been split off and the impact of the separate contributions to the
weak corrections have been studied, yielding results similar to those
for $\Pep\Pem\to\Pf\bar\Pf$.

\section{General result for one-loop electroweak logarithms}
\label{se:general-recipe}

A different method for the evaluation of the one-loop
logarithmic corrections has been developed in
\citeres{Denner:2001jv,Denner:2001gw,Pozzorini:2001}. This approach is
not related to individual Feynman diagrams but makes use of the
universality of the logarithmic corrections.

The derivation is carried out in the full spontaneously broken
electroweak Standard Model. It uses the Feynman gauge and dimensional
regularization with the regularization scale $\mu$ chosen as
$\mu^2=s$. In this setup the logarithms related to the running of the
couplings are obtained from the corresponding counter terms. The
double-logarithmic contributions originate from those one-loop
diagrams where soft--collinear gauge bosons are exchanged between
pairs of external legs. These are evaluated using the eikonal
approximation. The single-logarithmic corrections not related to
parameter renormalization originate from the wave-function
renormalization of the external particles and from diagrams with
collinear gauge bosons attached to external lines. The latter are
extracted in the collinear limit by means of Ward identities and are
found to factorize the Born amplitude.

The method is applicable to non-mass-suppressed exclusive processes
with arbitrary external states, including transverse and longitudinal
gauge bosons as well as Higgs fields.  Above the electroweak scale,
the photon, $\PZ$-boson and $\PW$-boson loops are treated in a
symmetric way, rather than split into electromagnetic and weak parts.
To this end, the logarithms originating from the electromagnetic loops
are split Ted into two parts: the contributions of a fictitious heavy
photon with mass $\MW$ are added to the $\PW$-boson and $\PZ$-boson
loops resulting in the ``symmetric electroweak'' contribution, and the
remaining part originating from the difference between the photon mass
and the mass of the $\PW$-boson is denoted as ``pure electromagnetic''
contribution.

The results can be summarized as follows: the universal,
angular-independent part of the double logarithms and the single
logarithms originating from collinear diagrams and wave-function
renormalization factorize the lowest-order matrix elements. The mixing
of photon and \PZ~boson entails, however, a mixing of the
corresponding matrix elements. At $1\TeV$, the corresponding relative
corrections to cross sections range between $-0.1\%$ and $-10\%$ per
external particle, the largest effects resulting from Higgs and gauge
bosons.  The angular-dependent logarithms, which originate from the
soft--collinear region, can be associated to pairs of external lines.
They do not obey a simple factorization, but require in addition
matrix elements where the external particles of the original process
are replaced by the corresponding $\SU(2)$ partners. Consequently, the
numerical size of these corrections depends strongly on the involved
matrix elements. External longitudinal gauge bosons have to be
replaced by the corresponding would-be Goldstone bosons.

Recently, the method of
\citeres{Denner:2001jv,Denner:2001gw,Pozzorini:2001} has been used to
study the effect of one-loop electroweak corrections on $\PW\PZ$ and
$\PW\ga$ production processes at the LHC \cite{Accomando:2001fn}.  It
was found that in the physically interesting region of large
transverse momentum and small rapidity separation of the gauge bosons
the corrections can be calculated in the leading-pole approximation.
As a result, the electroweak corrections lower the theoretical
predictions by 5--20\%


\section{Explicit two-loop calculations}
\label{se:two-loop}

To date, explicit two-loop calculations have essentially been
performed in order to check the reliability of the resummation
techniques and are restricted to the leading two-loop terms
[$\sim\al^2\log^4(s/\MW^2)$].

The corrections to the decay of an $\SUtwo\times\Uone$ singlet into
massless fermions have been calculated for the right-handed (abelian)
case by Melles \cite{Melles:2000ed} and for the left-handed
(non-abelian) case by Hori \etal\ \cite{Hori:2000tm}.

Beenakker and Werthenbach have developed a Coulomb gauge fixing for
massive gauge bosons \cite{Beenakker:2000na} that permits to isolate
the leading logarithms into self-energy diagrams. Explicit two-loop
results for the process $\Pep\Pem\to\Pf\bar\Pf$ have been given in
\citere{Beenakker:2000kb}. Recently, the calculation of the leading
two-loop logarithms for arbitrary external particles, \ie fermions,
longitudinal gauge bosons, Higgs bosons, and transverse gauge bosons
has been completed \cite{Beenakker:2001}. All these results are in
agreement with the exponentiation found in \citere{Fadin:2000bq}.

\section{Resummation of higher-order contributions}
\label{se:resummation}

The numerical size of the double logarithms at one loop suggests that
the corresponding leading two-loop corrections can be at the level of
$10\%$ at $1\TeV$. This calls for a resummation of these corrections.
This resummation has mainly been studied by applying QCD resummation
techniques to the electroweak theory, thereby assuming that at high
energies the symmetric phase of the electroweak theory can be used.
Leading [$\sim\al^n\log^{2n}(s/\MW)]$, sub-leading
[$\sim\al^n\log^{2n-1}(s/\MW)]$, and recently also sub-sub-leading
[$\sim\al^n\log^{2n-2}(s/\MW)]$ contributions have been studied.
 
Fadin \etal\ \cite{Fadin:2000bq} have resummed the leading
contributions by means of the infrared evolution equation for
non-radiative processes and for processes including soft gauge-boson
emission.  In all cases the Sudakov double logarithms were found to
exponentiate.  A resummation of Ciafaloni and Comelli
\cite{Ciafaloni:2000ub} based on soft gauge-boson insertions and
strong energy ordering gave different results. The explicit
calculations mentioned in \refse{se:two-loop} support the
exponentiation found in \citere{Fadin:2000bq}.

K\"uhn \etal\ have considered the process $\Pep\Pem\to\Pf\bar\Pf$ in
the limit of massless fermions and have used evolution equations to
resum all logarithmic corrections including the angular-dependent
ones. The leading and sub-leading contributions have been evaluated in
\citere{Kuhn:2000nn}. Recently also the sub-sub-leading contributions
(together with the constant one-loop terms) have been published
\cite{Kuhn:2001hz}. At TeV energies large cancellations between
leading, sub-leading and sub-sub-leading two-loop contributions have been
observed, while each of these contributions can reach several per cent
at $1\TeV$.

Using the concept of splitting functions, Melles has extended the
infrared-evolution equation approach to the sub-leading level
considering processes involving fermions and transverse gauge bosons
\cite{Melles:2001gw}. In \citere{Melles:2001ia} he applied this method
to processes involving external longitudinal gauge bosons and Higgs
bosons, where additional logarithms related to Yukawa couplings
appear. The effect of the running gauge couplings at the sub-leading
level has been discussed in \citere{Melles:2001mr}.  A review of this
work can be found in \citere{Melles:2001ye}, where it is shown that
the sub-leading two-loop logarithms can yield corrections of up to
4\% at $1\TeV$.  Very recently a generalization including also the
angular-dependent logarithms for arbitrary processes has been proposed
\cite{Melles:2001dh}.

\section{Non-cancellation of leading electroweak logarithms in inclusive
  quantities}
\label{se:non-cancellation}

As discussed above, virtual one-loop corrections contain double
logarithms which arise from the soft--collinear limit. In QED and QCD,
these double logarithms cancel when adding the corresponding real
corrections and, in case of QCD, averaging over the colours of the
initial state particles, such that inclusive quantities are free of
double logarithms.  It has been pointed out in
\citere{Ciafaloni:2000df}, that the situation is different in the
electroweak Standard Model. Because the initial states
($\Pep\Pem,\Pq\Pq',\ga\ga$, \ldots) carry a definite non-abelian
flavour, the double logarithms do not cancel in inclusive hard cross
\looseness -1
sections and the Bloch--Nordsieck theorem is violated.

The origin of this non-cancellation can be understood by considering,
for instance, fully inclusive $\Pep\Pem$ annihilation: virtual
corrections simply multiply the lowest-order matrix elements and are
proportional to the lowest-order cross section $\si_{\Pep\Pem}$. The
same holds for the corrections related to real emission of a photon or
a \PZ~boson. However, the real emission of a charged \PW~boson changes
the isospin of the incoming electron and turns it into a neutrino. The
corresponding double-logarithmic corrections are proportional to the
cross section $\si_{\Pep\Pne}$ and do not cancel the corresponding
virtual corrections.  In fact, the resulting electroweak corrections
exceed the QCD corrections in the TeV range.

The Bloch--Nordsieck-violating double logarithms can be exponentiated
\cite{Ciafaloni:2000rp} and lead to suppression of cross-section
differences within weak-isospin multiplets. Recently, it has been
found that Bloch--Nordsieck violation can also occur in spontaneously
broken abelian gauge theories, if the incoming particles are mass
eigenstates that do not coincide with gauge eigenstates
\cite{Ciafaloni:2001vt}. In the Standard Model this mechanism becomes
relevant for incoming longitudinal gauge bosons or Higgs bosons
\cite{Ciafaloni:2001vu}, since these can turn into each other via the
\looseness -1
emission of a \PZ~ boson.

\section{Conclusions}
\label{se:conclusion}

At energies above the electroweak scale electroweak logarithms can
lead to large corrections of several tens of per cent.  Unlike in QED
and QCD, the leading double-logarithmic corrections do not cancel in
inclusive quantities. On the other hand, at TeV energies leading,
sub-leading, and sub-sub-leading virtual corrections tend to cancel
each other, and the size of the total corrections depends strongly on
the energy and on the observable under consideration. It is clear that
these corrections must be under control in order to analyse
experiments at future colliders.

The virtual electroweak logarithmic corrections at one loop have been
calculated for various processes, and a general method for the
evaluation of these corrections has been developed. The leading
double-logarithmic corrections can be resummed via exponentiation of
the one-loop expression.  First results for the sub-leading and
sub-sub-leading corrections at two loops exist. The investigation of
these corrections requires further studies.

\acknowledgments
I would like to thank S. Pozzorini for help in the preparation of the
manuscript. 



\end{document}